\documentclass[reviewcopy]{elsart}
\usepackage{graphics,natbib,graphicx,psfig,amssymb,ccaption}
\journal{New Astronomy}
\begin{document}
\begin{frontmatter}
\title{The early-type near-contact binary system V337 Aql revisited}
\author[canakkale]{M. T\"uys\"uz\corauthref{cor}},
\corauth[cor]{Corresponding author. Fax: +90 286 218 05 33}
\ead{mehmettuysuz@comu.edu.tr}
\author[canakkale]{F. Soydugan},
\author[istanbul]{S. Bilir},
\author[canakkale]{E. Soydugan},
\author[canakkale]{T. \c{S}eny\"uz},
\author[istanbul]{T. Yontan}
\address[canakkale]{\c Canakkale Onsekiz Mart University, Faculty of Sciences
and Arts, Physics Department, 17100 \c Canakkale, Turkey}
\address[istanbul]{Istanbul University, Faculty of Sciences, Department of Astronomy
and Space Sciences, 34119 University, Istanbul, Turkey}

\begin{abstract}
The close binary V337 Aql consists of two early B-type components
with an orbital period of 2.7339 d. New multi-band photometric
observations of the system together with published radial velocities
enabled us to derive the absolute parameters of the components. The
simultaneous light and radial velocity curves solution yields masses
and radii of \emph{M$_{1}$}=17.44$\pm$0.31 \emph{M$_\odot$} and
\emph{R$_{1}$}=9.86$\pm$0.06 \emph{R$_\odot$} for the primary and
\emph{M$_{2}$}=7.83$\pm$0.18 \emph{M$_\odot$} and
\emph{R$_{2}$}=7.48$\pm$0.04 \emph{R$_\odot$} for the secondary
component. Derived fundamental parameters allow us to calculate the
photometric distance as 1355$\pm$160 pc. The present analysis
indicates that the system is a near-contact semi-detached binary, in
which a primary star is inside its Roche lobe with a filling ratio
of 92 percent and the secondary star fills its Roche lobe. From
\emph{O-C} data analysis, an orbital period decrease was determined
with a rate of -7.6 $\times$ 10$^{-8}$ yr$^{-1}$. 
Kinematic analysis reveals that V337 Aql has
a circular orbit in the Galaxy and belongs to a young thin-disc
population.

\end{abstract}

\begin{keyword}
techniques: photometric - stars: binaries: eclipsing - stars: individual: V337 Aquilae
\end{keyword}

\end{frontmatter}

\section{Introduction}
V337 Aql (HD 177284, BD -02 4840, SAO 142979, V = 8$^m$.64) is a
$\beta$ Lyr type eclipsing binary system with an orbital period of
$\sim$ 2.7339 days. The first estimation of the photometric
parameters of the system based on its photographic light curve
analysis was given by Wright $\&$ Dugan (1936). Using several plate
spectra, Feast \& Thackeray (1963) measured the radial velocities
and classified the system as a double-line spectroscopic binary. In
the study of Catalano et al. (1971), the first complete
photoelectric light curve of the system was published and reported
asymmetries and variations on its light curve. They also announced
that the orbital period of the system was decreasing. Alduseva
(1977) mentioned the amplitude variations and instabilities on the
\emph{UBV} light curves of V337 Aql as being connected with mass
exchange between the components and also, using radial
velocity measurements, estimated the masses of the
components to be \emph{M$_{1}$} $\sim$ 16\emph{M$_\odot$} and
\emph{M$_{2}$} $\sim$ 10\emph{M$_\odot$}. \emph{B} and \emph{V}
light curves obtained by Alduseva (1977) were also solved by
Giuricin $\&$ Mardirossian (1981); however, their results strongly
differed from those of Catalano et al. (1971). Mayer (1987) reported
the changes in the orbital period related with the light-time
effect. A study on the orbital period variation of V337 Aql was
published by \u{S}imon (1999), in which, although the data in the
\emph{O-C} diagram indicated scattering, he suggested a possible
decrease in the orbital period, which corresponded to
\emph{$\Delta$P/P} $\approx$ -6.5$\times$10$^{-10}$ days$^{-1}$. Lastly,
Mayer et al. (2002) measured the radial velocities of the components
of V337 Aql based on high resolution echelle spectra and gave the
orbital parameters. They also analyzed the light curve published by
Catalano et al. (1971) and reported the absolute parameters of the
components.

In this study, new photometric analysis of the early-type eclipsing
binary system V337 Aql is presented. After presenting information
and background studies of the system, new photometric data are
described in Section 2, followed by analysis of the orbital period
variation. In the next section, the simultaneous solution of
\emph{BVR} light curves together with published radial velocity
curves is outlined. In the last section, the results and discussion,
which include absolute parameters of the components, comparing the
system with the similar eclipsing binaries and also kinematic
properties are presented.

\section{Observations}
New photometric observations of V337 Aql were performed at the
\c{C}anakkale Onsekiz Mart University Observatory over 22 nights in
the observation season of 2012. During the observations of the
system, a 60-cm Cassegrain telescope, equipped with Apogee ALTA U42
CCD camera and Bessell \emph{BVR} filters, was used. The camera is
of 13.5 $\times$ 13.5 microns pixel size and provides an effective
field of view (FOV) of 17$^{'}$.5  $\times$ 17$^{'}$.5. The total
number of the points obtained was 9880 in all filters. TYC
5132-878-1 and TYC 5128-840-1 were used as comparison and check
stars, respectively. Reductions of the CCD images and aperture
photometry were made using the C-Munipack
\footnote[1]{http://c-munipack.sourceforge.net/} software package
in the standard mode. The difference magnitude between the
comparison and check stars did not indicate any significant light
variation. The errors of individual observational points were
estimated as about 0$^{m}$.01 in all the filters. Figure 2
represents the \emph{BVR} light curves of the system. The orbital
phase in the figure was computed according to the following light
elements given by Kreiner (2004):

\begin{equation}
HJD~(Min I) = 2452500.334 + 2^{\rm d}.733882(1)\times E
\end{equation}

All observational data shown in Fig. 2 was used the light curve
synthesis given in Section 4.

\section{Orbital Period Variation}

The \emph{O-C} diagram of V337 Aql in the database of the \emph{O-C}
Atlas of Eclipsing Binaries (Kreiner et al. 2001) indicates orbital
period change. Although the visual (v) and photovisual (p) minima
times contain relatively large errors and also large scatter, the
orbital period variation can be studied using photographic (pg),
photoelectric (pe) and CCD minima times. We collected the data from
the \emph{O-C} Atlas of Eclipsing Binaries (Kreiner et al. 2001) and
combined them with three minima times (one primary and two
secondary) determined from the new data listed in Table 1. New
minima times were calculated using the Kwee $\&$ van Woerden (1956)
method. For the \emph{O-C} analysis, 20 photographic, 8
photoelectric and 3 CCD minima times were used, while their weights
were selected as per the following: pg = 5, pe and CCD = 10. The
\emph{O-C} values were plotted against the epoch number in Fig. 1.
As shown in Fig. 1, the general trend of \emph{O-C} data can be
represented by a downward parabola indicating a long-term orbital
period decrease. Hence, by applying a weighted least-squares
fitting, the following quadratic ephemeris with errors is derived:

\begin{equation}
HJD~(Min I) = 2434896.8712(7) + 2^{\rm d}.7338836(1)\times E -7.8(2)\times10^{-10}\times E^{2}
\end{equation}

Using the quadratic term, the rate of orbital period decrease can be
derived to be \emph{dP/dt} = -7.6 $\times$ 10$^{-8}$ yr$^{-1}$.
As can be seen in Fig. 1, the downward parabolic term represents the
general trend of the \emph{O-C} data well and the last three minima
times obtained in this study also support the variation.

\begin{table*}
  \begin{center}
  \caption{Minima times obtained in this study for V337~Aql.\vspace{3 mm}}\label{tbl-1}
    \begin{tabular}{lccc}
  \hline \hline
HJD+2400000  & Error  & Type & Filters   \\
\hline
56114.4964  & $\pm$ 0.0004       & Primary   & \emph{BVR}        \\
56110.3951  & $\pm$ 0.0005       & Secondary  & \emph{BVR}\\
56140.4627  & $\pm$ 0.0004       & Secondary  & \emph{BVR}\\
\hline
\end{tabular}
\end{center}
\vspace{4 mm}
\end{table*}

\begin{figure}
\begin{center}
\includegraphics[width=96mm,height=90mm]{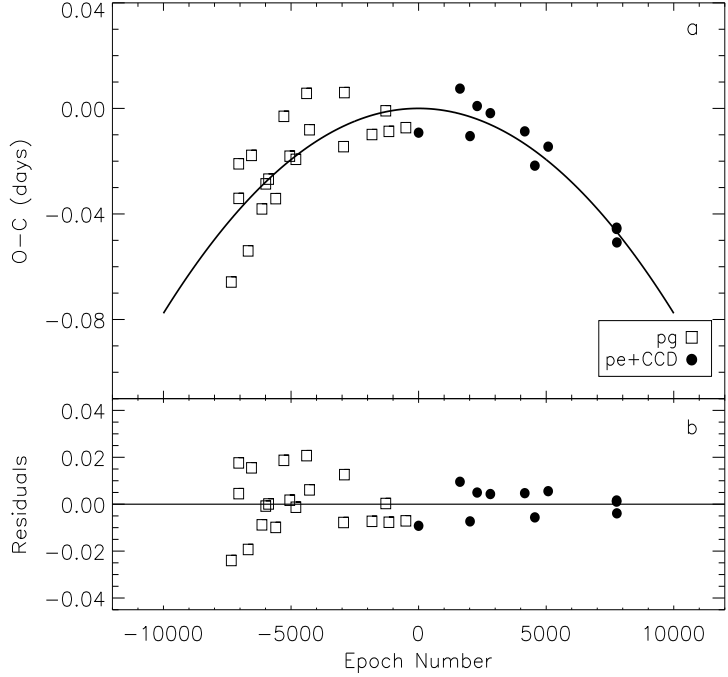}
\caption{$O$--$C$ diagram of V337 Aql showing (a) downward parabolic
representation (solid line) of $O$-$C$ changes in the system and (b)
residuals from the best theoretical curve (b).\vspace{4 mm}} \label{fig1}
\end{center}
\end{figure}

\section{Light Curve Synthesis}
The \emph{BVR} light curves obtained in this study and the radial
velocities given by Mayer et al. (2002) of V337 Aql were solved
simultaneously using the Wilson-Devinney (W–D) program (Wilson \&
Devinney, 1973). The difference magnitudes (867 points in \emph{B},
1152 points in \emph{V} and 957 points in \emph{R} filters) for each
filter were converted to intensity units using the mean magnitudes
at 0.25 orbital phase. Although a semi-detached configuration was
suggested for V337 Aql (Catalano et al. 1971, Alduseva 1977,
Giuricin and Mardirossian 1981, Mayer et al. 2002), firstly, the
detached mode (Mode 2 in W-D code) was used to check whether the
surface potentials of the components reached their Roche limits or
not. After trials, it was seen that the secondary component filled
its Roche lobe and it was concluded that the semi-detached system
configuration should be taken into account. Therefore, Mode 5 in 
W-D code was selected for the analysis.

In the light curve solutions, some parameters were fixed as the following:
The effective temperature of the primary component was adopted as 28000 K 
(Popper 1980) for the B0.5 spectral class suggested by Roman (1956) and 
Mayer et al. (2002). The bolometric albedos
\emph{A$_{1,2}$} were taken from Rucinski (1969) to be 1.0 for the
components with radiative envelope. The bolometric gravity-darkening
coefficients \emph{g$_{1,2}$} were set to 1.0 for radiative
atmospheres from von Zeipel (1924). The corresponding logarithmic
bolometric and monochromatic limb-darkening coefficients used were
from van Hamme's (1993) tables. The secondary minimum is seen at 0.5
phase and the ascent and descent duration of the secondary and
primary minimum are the same, therefore, a circular orbit (\emph{e
}= 0) was assumed. The third light (\emph{$l_{3}$} = 0) was
fixed to zero since no evidence was found during trials. The adjusted 
parameters are the semi-major axis of orbit (\emph{a}), radial velocity 
of the system's mass center ($V_{\gamma}$), phase shift ($\phi$), orbital
inclination (\textit{i}), surface temperature of the secondary
component ($T_{2}$), non-dimensional surface potential of primary
component ($\Omega_{1}$), mass ratio (\textit{q}) and the fractional
luminosity of the primary component ($L_{1}$). The input values of
\emph{a}, $V_{\gamma}$ and \textit{q} were taken into account from
the orbit solution of Mayer et al. (2002).

Results from our simultaneous solution of \emph{BVR} light curves
are given in Table 2. The corresponding theoretical light curves are
plotted in Fig. 2 overlaying the observational points. Roche
geometry of the system is represented by Binary Maker code
(Bradstreet 1990), as illustrated in Fig. 3.

\begin{table}
  \begin{center}
  \caption{Parameters of V337 Aql derived from simultaneous analysis of \emph{BVR} light curves and RVs of components. Probable errors in last digits are given in parenthesis.\vspace{3 mm}}\label{tbl-3}
    \begin{tabular}{llll}
  \hline\hline
 Parameter                    & W-D solution  & \,\,\,\,\,\,   Parameter                &\,\,\,    W-D solution                        \\
\hline
 \textit{a}(R$_{\bigodot}$)   & 24.12(11)     & \,\,\,\,\,\, \textit{L}$_{1}$ /( \emph{L$_{1}$ + L$_{2}$})- \emph{B} &\,\,\, 0.708(15)   \\
 \textit{i}(deg)            & 83.66(8)        &\,\,\,\,\,\,\, \textit{L}$_{1}$ /( \emph{L$_{1}$ + L$_{2}$})- \emph{V} & \,\,\, 0.701(14)   \\
 \textit{V$_{\gamma}$}(km s$^{-1}$)& 38.6(9)  &\,\,\,\,\,\,\, \textit{L}$_{1}$ /( \emph{L$_{1}$ + L$_{2}$})- \emph{R} &\,\,\, 0.696(14)   \\
 \textit{T}$_{1}$(K)          & 28000$^{a}$   &\,\,\,\,\,\,\, \textit{L}$_{2}$ /( \emph{L$_{1}$ + L$_{2}$})- \emph{B} &\,\,\, 0.292  \\
 \textit{T}$_{2}$(K)          & 23640(40)     &\,\,\,\,\,\,\, \textit{L}$_{2}$ /( \emph{L$_{1}$ + L$_{2}$})- \emph{V} &\,\,\, 0.299   \\
 $\Omega$$_{1}$               & 2.9803(54)    &\,\,\,\,\,\,\, \textit{L}$_{2}$ /( \emph{L$_{1}$ + L$_{2}$})- \emph{R} &\,\,\, 0.304   \\
 $\Omega$$_{2}$               & 2.7772        &\,\,\,\,\,\,\, \textit{r}$_{1}$ (mean)      &\,\,\, 0.409(5)         \\
 Phase shift                  & 0.0007(3)     &\,\,\,\,\,\,\, \textit{r}$_{2}$ (mean)      &\,\,\, 0.310(4)        \\
 \textit{q} (=\emph{M$_{2}$/M$_{1}$})& 0.449(5)& &       \\
 \hline
\end{tabular} \\
\footnotesize $^{a}$: Adopted from Popper (1980)\vspace{5 mm}
\end{center}
\end{table}

\begin{figure}
\begin{center}
\includegraphics[width=110mm,height=70mm]{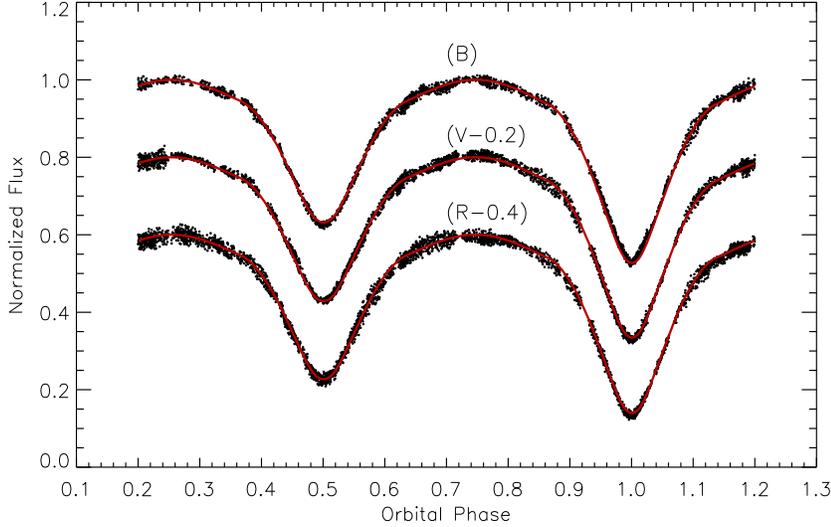}
\caption{Observed and theoretical light curves of V337 Aql in \emph{BVR} filters.}
\end{center}
\label{fig:lc_V337 Aql}
\vspace{4 mm}
\end{figure}

\begin{figure}
\begin{center}
\includegraphics[width=60mm,height=27mm]{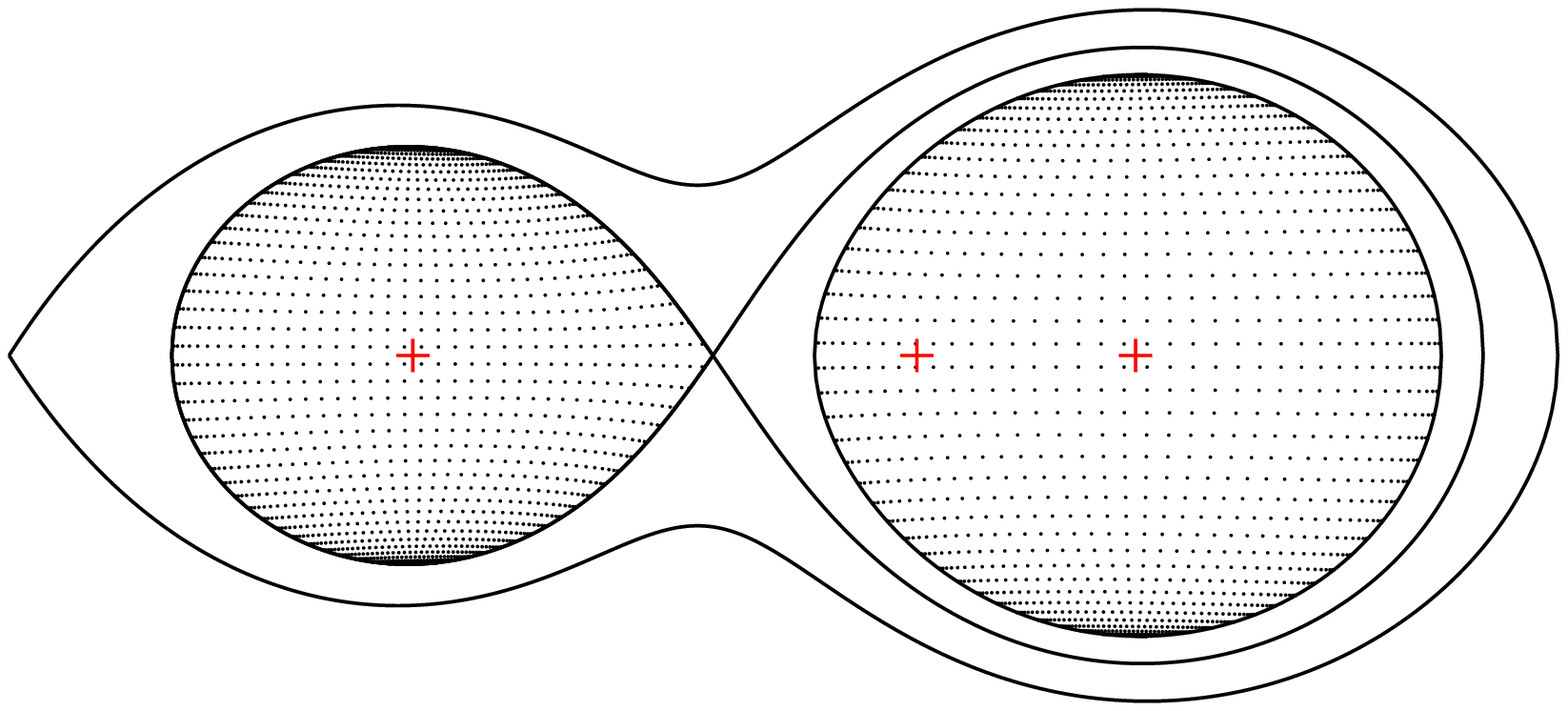}
\caption{Roche geometry of V337 Aql at orbital phase 0.75.}
\label{fig3}
\end{center}
\vspace{4 mm}
\end{figure}

\section{Results and Discussion}

The new \emph{BVR} photometric observations of the early-type
eclipsing binary V337 Aql combined with radial velocities from the
literature (Mayer et al. 2002) were analyzed to determine its
fundamental properties. The system configuration was found to be 
semi-detached; the primary component being
in its Roche lobe, while the secondary component filling its Roche
lobe. V337 Aql can be classified as a near-contact binary since the
Roche lobe filling ratio of the hotter component is about 92
percent. Based on the photometric and spectroscopic elements in
Table 2, the basic astrophysical properties of the components and
the orbital parameters of the system were derived and given Table 3.
For the calculations, solar values and bolometric corrections (\emph{BC}) for
the components were adopted from Drilling \& Landolt (2000).
The masses of the components were calculated as 17.44 M$_\odot$ and 7.83
M$_\odot$ for the primary and secondary, respectively, which are
consistent with the masses of B0 and B3 main sequence stars
according to the tables of Drilling \& Landolt (2000). As mentioned
by Mayer et al. (2002), the radii of the components have higher
values compared with main-sequence stars. This is probably related
the effects of mass and angular momentum transfer and also mass loss
from the system. This scenario may be expected since the system is
semi-detached and has very hot components.

Using the \emph{U-B} and \emph{B-V} colours of the system given by
Popper \& Dumont (1977), we applied the \emph{Q}-method given by
Johnson \& Morgan (1953) to determine colour excess \emph{E(B-V)},
which was found to be 0.80 mag. The high value of colour excess is
consistent with the location of the system in the Milky Way
(Galactic coordinates: \emph{l}= 32.57 deg and \emph{b}=-03.76 deg).
Interstellar absorption in the \emph{V} filter was calculated to be
2.48 mag based on the common formula \emph{A$_{v}=3.1\times
E(B-V)$}. Using apparent system magnitude, interstellar
extinction and light ratio of components, the photometric distance
of the system was estimated as 1355$\pm$160 pc.

\begin{table*}
\begin{center}
\caption{Fundamental parameters of close binary V337 Aql parameters of V337~Aql. Uncertainties
are given in brackets.} \label{astparams}
\vspace{5 mm}
\renewcommand{\arraystretch}{1}
\renewcommand{\tabcolsep}{0.1cm}
\begin{tabular}{lccc}\hline\hline
Parameter                          & Symbol         & Primary    & Secondary
\\
\hline
Spectral type                      & Sp             &  B0 V      &    B3 V \\
Mass (M$_\odot$)                   & \emph{M}       & 17.44(0.31)   &   7.83 (0.18)\\
Radius (R$_\odot$)                 & \emph{R}       & 9.86(0.06) &  7.48 (0.04)\\
Separation (R$_\odot$)             & \emph{a}  & \multicolumn{2}{c}{24.12 (0.11)}\\
Orbital period (days)              & \emph{P}  & \multicolumn{2}{c}{2.7338794} \\
Orbital inclination ($^{\circ}$)   & \emph{i}       & \multicolumn{2}{c}{83.66(0.08)}    \\
Mass ratio                         & \emph{q}       & \multicolumn{2}{c}{0.449(0.005)} \\
Eccentricity                       & \emph{e}       &\multicolumn{2}{c}{0.0}    \\
Surface gravity (cgs)              & $\log g$       & 3.69(0.01)& 3.58(0.01)       \\
Integrated visual magnitude (mag)          & \emph{V}       & \multicolumn{2}{c}{8.64*}   \\
Integrated colour indices (mag)            & $U-B$, $B-V$      & \multicolumn{2}{c}{--0.52*, 0.49*} \\
Colour excess (mag)                &$E(B-V)$        & \multicolumn{2}{c}{0.80}   \\
Visual absorption (mag)            & $A_{\rm v}$    & \multicolumn{2}{c}{2.48}   \\
Intrinsic colour index (mag)       & $(B-V)_{\rm0}$ & \multicolumn{2}{c}{-0.31} \\
Temperature (K)                    & $T_{\rm eff}$  & 28000 (500)   & 23640(500)   \\
Luminosity (L$_\odot$)             & $\log$ \emph{L}& 4.73 (0.04)   & 4.20(0.04)   \\
Bolometric magnitude (mag)         &$M_{\rm bol}$   & --7.08 (0.09) & --5.75(0.10) \\
Abs. visual magnitude (mag)        &$M_{\rm v}$     & --4.11(0.04)   & --3.20(0.04)\\
Bolometric correction (mag)        &\emph{BC}       & --2.97  & --2.55 \\
Systemic velocity (km\,s$^{-1}$)   &$V_{\gamma}$    & \multicolumn{2}{c}{38.6(0.9)}\\
Distance (pc)                      &\emph{d}        & \multicolumn{2}{c}{1355(160)}  \\
Proper motion (mas yr$^{-1}$) &$\mu_\alpha cos\delta$, $\mu_\delta$ &\multicolumn{2}{c}{-0.70(0.80), -4.30(0.80)**}  \\
Space velocities (km\,s$^{-1}$) & $U, V, W$ & \multicolumn{2}{c}{35.02 (3.33), 8.60 (5.16), -4.60 (5.23)} \\
\hline
\end{tabular}
{*Popper \& Dumont (1977), ** Zacharias et al. (2013).}
\end{center}
\end{table*}

The locations of the components of V337 Aql and also some early-type 
eclipsing binaries showing $\beta$ Lyr type light curves (Table 4) in the
Hertzsprung-Russell (HR) diagram are presented in Fig. 4. Zero Age
Main Sequence (ZAMS), Terminal Age Main Sequence (TAMS) and all
evolutionary tracks and isochrones for solar chemical composition
are taken from Ekstr\"{o}m et al. (2012). The surface gravity values
of the components of V337 Aql in Table 3 and the locations of the
components in the HR diagram indicate that the component stars are
within the main-sequence band and close to TAMS. The location of the
primary component of V337 Aql in the diagram is very close to the
evolutionary track of 17 M$_\odot$ calculated for a single star with
solar abundance, which is consistent with its mass value
and an isochrone of about 8 Myr. On the other hand, the
less massive component seems to have higher luminosity and
temperature with respect to its mass. The secondary
component appears close to an isochrone of about 15 Myr. 

\begin{table}
\caption{Basic physical parameters of some early-type eclipsing
binaries indicating $\beta$ Lyr type light curves.} \label{table1}
    $$
    \renewcommand{\arraystretch}{1}
        \begin{tabular}{lcccccccc}
    \hline\hline
System     &  Spectral   & System     & P  & \emph{$M_{1}$} (M$_\odot$) & \emph{$R_{1}$} (R$_\odot$)& $T_{1}$ (K) & References \\
           &  Type       &Type$^{a}$& (day)           & \emph{$M_{2}$} (M$_\odot$) & \emph{$R_{2}$} (R$_\odot$)& $T_{2}$ (K) &  \\
\hline
CC Cas     &  O8.5III    & D          & 3.366344        & 18.3                      & 10.08            & 34500 &1  \\
           &  B0.5       &            &                 & 7.6                       & 4.02             & 28300 &\\
TU Mus     &  O7V        & C          & 1.387283        & 16.7                      & 7.2              & 38700 &2 \\
           &  O8V        &            &                 & 10.4                      & 5.7              & 33200 & \\
XZ Cep     &  O9.5V      & D          & 5.097253        & 15.8                      & 7.0              & 30000 &3 \\
           &  B1III      &            &                 & 6.4                       & 10.5             & 23120 &\\
V606 Cen   &  B0-0.5V    & C          & 1.4950996       & 14.7                      & 6.8              & 29200 &4  \\
           &  B2-3V      &            &                 & 8.0                       & 5.2              & 21870 &\\
IU Aur     &  B0.5       & SD         & 1.811474        & 14.5                      & 6.2              & 29825 &5  \\
           &  B0.5       &            &                 & 7.3                       & 5.7              & 26830 &\\
   \hline
\end{tabular}
$$
\begin{small}
$^{a}$ D: detached, SD: semi-detached, C: contact systems; References: (1) Hill et al. (1994), (2) Penny et al. (2008), (3) Harries et al. (1997), (4) Lorenz et al. (1999), (5) Harries et al. (1998) \\
\end{small}
\end{table}

When we look at the positions of the less massive components of early-type
binaries in the HR diagram, which were determined using the
parameters listed in Table 4, all components have higher luminosity
and temperature values with respect to their masses, similar to the
cooler component of V337 Aql. This phenomenon is very common for the
cooler components of classical Algols (\.{I}bano\={g}lu et al.
2006). The primaries of the massive binaries XZ Cep, V606 Cen and IU
Aur seem to be close to the evolutionary tracks calculated for their
masses. However, the more massive components of TU Mus and CC Cas
are far from their expected positions in the HR diagram, estimated
using the single-star evolutionary models.
These discrepancies in V337 Aql and the other examples given
may be related the effects of mass-transfer and/or loss during
evolution of the systems.

\begin{figure}
\begin{center}
\includegraphics[width=100mm,height=87mm]{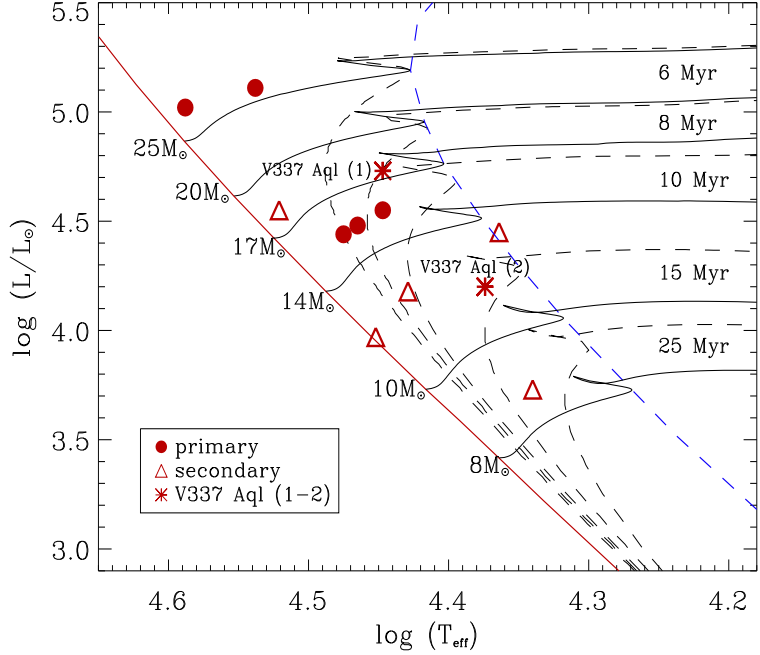}
\caption{Locations of primary and secondary components of some
early-type eclipsing binaries (in Table 4) together with components
of V337 Aql in log \emph{L} - log \emph{T$_{eff}$} plane. The
evolutionary tracks for different masses in solar unit (solid
lines), isochrones for different ages in Myr (dashed lines),
ZAMS and TAMS for solar chemical composition are adopted from
Ekstr\"{o}m et al. (2012).\vspace{5 mm}} \label{fig4}
\end{center}
\end{figure}

The changing character of the orbital period of V337 Aql is
monotonic, similar to several early-type eclipsing binaries (e.g. RY
Sct, Z Vul, BF Aur; \u{S}imon 1999). The orbital period of V337 Aql
has been decreasing at a rate of 1.8 s century$^{-1}$. Although the
system is semi-detached and an orbital period increase may be
expected from the less massive star to the more massive one as in
classical Algols, the orbital period of V337 Aql is decreasing. 
This may be interpreted as mass and angular momentum loss
from the system resulting from radiative forces, which may be
produced by the hot components. One of the rare examples indicating
orbital period decrease among early-type eclipsing binaries is RY
Sct. Mass loss from the system at a rate of 6$\times$10$^{-5}$
M$_\odot$ yr$^{-1}$, which may be used to explain the decrease in
the orbital period, was determined using radio observations by
Milano et al. (1981).

In order to obtain the kinematical properties of V337~Aql, we used
the system's center of mass' velocity, distance and proper motion
values, as given in Table 3. The proper motion data were taken from
the Fourth US Naval Observatory CCD Astrograph Catalog
(UCAC4; Zacharias et al. 2013), whereas the center of mass' velocity
and distance were obtained from this study. The system's space
velocity was calculated using the algorithm given by Johnson \& Soderblom (1987).
The $U$, $V$ and $W$ space velocity components and their errors were obtained
and are given in Table 3. To obtain the precise space velocity the
first-order galactic differential rotation correction was taken into
account (Mihalas \& Binney 1981), and 19.80 and 2.73 kms$^{-1}$
differential corrections were applied to the $U$ and $V$ space
velocity components, respectively. The $W$ velocity is not affected
in this first-order approximation. As for the Local Standard Rest
(LSR) correction, Co\c{s}kuno\v{g}lu et al. (2011)'s values (8.5, 13.38,
6.49)$_{\odot}$ kms$^{-1}$ for thin disc stars were used and the
final space velocity of V337~Aql was obtained as $S=36.35\pm 8.07$
kms$^{-1}$. This value is in agreement with the space velocities of
other young stars.

To determine the population type of V337~Aql the Galactic orbit of
the system was examined. Using Dinescu et al. (1999) N-body code, the
system's apogalactic ($R_{max}$) and perigalactic ($R_{min}$)
distances were calculated to be 8.41 and 6.58 kpc, respectively. In
addition, the maximum possible vertical separation from the Galactic
plane is $|z_{max}|$=0.11 kpc for the system. The following formulae
were used to derive the planar ($e_p$) and vertical ($e_v$)
ellipticities:

\begin{equation}
e_p=\frac{R_{max}-R_{min}}{R_{max}+R_{min}},
\end{equation}

\begin{eqnarray}
e_v=\frac{(|z_{max}|+|z_{min}|)}{R_m},
\end{eqnarray}
where $R_m=(R_{max}+R_{min})/2$ (Pauli 2005). Due to
$z$-excursions $R_p$ and $R_a$ may vary, however this variation is
not more than 5 per cent. The planar and vertical ellipticities were
calculated as $e_p=0.12$ and $e_v=0.03$. These values show that
V337~Aql is orbiting the Galaxy in a circular orbit, and that the
system belongs to an extremely young thin-disc population.

In conclusion, we might mention that in order to obtain evidence of
mass loss from the system and also membership of young-disc
population, further high resolution spectroscopic observation in
different wavelength regions for V337 Aql should be made. Moreover,
comparing early-type binaries with similar properties shows that
more detailed evolutionary models for early-type systems with mass
transfer and loss are needed in order to obtain more information
about evolutionary differences between these systems.

\section{Acknowledgments}

This research was supported by the Scientific and Technological
Research Council of Turkey (T\"UB\.ITAK, Grant no. 111T224) and 
Scientific Research Projects Coordination Unit of Istanbul University. 
Project number 3685. We thank \c{C}anakkale Onsekiz Mart 
University Astrophysics Research Center and Ulup{\i}nar Observatory 
together with \.{I}stanbul University Observatory Research and 
Application Center for their support and allowing use of IST60 
telescope. This research has made use of the SIMBAD, and NASA 
Astrophysics Data System Bibliographic Services.


\begin{thebibliography}{}

\bibitem[Alduseva (1977)]{alduseva77}
Alduseva, V.Y., 1977. PZ 20, 375

\bibitem[Bradstreet (1990)]{bradstreet90}
Bradstreet, D.H., 1990. BAAS 22, 1293

\bibitem[Catalano (1971)]{catalano71}
Catalano, F.A., Catalano, S., Rodon\'{o}, M., 1971. Ap\&SS 11, 232

\bibitem[Co{\c s}kuno{\v g}lu et.(2011)]{Coskunoglu11}
Co{\c s}kuno{\v g}lu, B., et al., 2011. MNRAS 412, 1237

\bibitem[Dinescu et al.(1999)Dinescu, Girard \& van Altena]{Dinescu99}
Dinescu, D. I., Girard, T. M., van Altena, W. F., 1999. AJ 117, 1792

\bibitem[Drilling and Landolt (2000)]{drilling00} Drilling, J. S., Landolt, A. U.,
2000. Allen's Astrophysical Quantities, 4th ed., Edited by Arthur N.
Cox. ISBN: 0-387-98746-0. Publisher: New York: AIP Press, Springer

\bibitem[Ekstrom et al. (2012)]{ekstrom12}
Ekstr\"{o}m, S., Georgy, C., Eggenberger, P., et al., 2012. A\&A
537, 146

\bibitem[Feast (1963)]{feast63}
Feast, M.W., Thackeray, A.D., 1963. MmRAS 68, 173

\bibitem[Giuricin (1981)]{giuricin81}
Giuricin, G., Mardirossian, F., 1981. A\&AS 45, 499

\bibitem[Harries et al. (1997)]{harries97}
Harries, T. J., Hilditch, R. W., Hill, G., 1997. MNRAS 285, 277

\bibitem[Harries et al. (1998)]{harries98}
Harries, T. J., Hilditch, R. W., Hill, G., 1998. MNRAS 295, 386

\bibitem[Hill et al. (1994)]{hill94}
Hill, G., Hilditch, R. W., Aikman, G. C. L., Khalesseh, B., 1994.
A\&A 282, 455

\bibitem[\.{I}bano\u{g}lu et al. (2006)]{ibanoglu06}
\.{I}bano\u{g}lu, C., Soydugan, F., Soydugan, E.,
Dervi\c{s}o\u{g}lu, A., 2006. MNRAS 373, 435

\bibitem[Johnson and Morgan (1953)]{morgan53}
Johnson, H. L., Morgan, W. W., 1953. ApJ 117, 313

\bibitem[Johnson \& Soderblom(1987)]{Johnson87}
Johnson, D. R. H., Soderblom, D. R., 1987. AJ 93, 864

\bibitem[Kreiner et al. (2001)]{kreiner01}
Kreiner,  J. M., Kim, C.-H., Nha, I.-S., 2001. An Atlas of
\emph{O-C} Diagrams of Eclipsing Binary Stars, Parts 1--6, Cracow:
Pedagogical University Press

\bibitem[Kreiner (2004)]{kreiner04}
Kreiner, J.M., 2004. AcA 54, 207

\bibitem[Kwee (1956)]{kwee56}
Kwee, K. K., van Woerden, H., 1956. BAN 12, 327

\bibitem[Lorenz et al. (1999)]{lorenz99}
Lorenz, R., Mayer, P., Drechsel, H., 1999. A\&A 345, 531

\bibitem[Mayer (1987)]{mayer87} Mayer, P., 1987. BAICz 38, 58

\bibitem[Mayer et al. (2002)]{mayer02} Mayer, P., Lorenz, R.,
Drechsel, H., 2002. A\&A 388, 268

\bibitem[Mihalas \& Binney(1981)]{Mihalas81}
Mihalas, D., Binney, J., 1981. Galactic Astronomy: Structure and
Kinematics, 2nd edition, San Francisco, CA, W. H. Freeman and Co.

\bibitem[Milano et al. (1981)]{milano81}
Milano, L., Vittone, A., Ciatti, F., Mammano, A., Margoni, R.
Strazzulla, G., 1981. A\&A 100, 59

\bibitem[Pauli(2005)]{Pauli05}
Pauli, E. M., 2005. Prof. G. Manev's Legacy in Contemporary
Astronomy, Theoretical and Gravitational Physics (Eds. V. Gerdjikov
and M. Tsvetkov), 185, Sofia, Bulgaria, Heron Press Limited, 2005.

\bibitem[Penny et al. (2008)]{penny08}
Penny, L. R., Ouzts, C., Gies, D. R., 2008. ApJ 681, 554

\bibitem[Popper and Dumont (1977)]{popper77} Popper, D. M., Dumont,
P., 1977. AJ

\bibitem[Popper (1980)]{popper80}
Popper, D.M., 1980. ARA\&A 18, 115

\bibitem[Roman (1956)]{roman56}
Roman, N.G., 1956. ApJ 123, 246

\bibitem[Rucinski (1969)]{rucinski69}
Rucinski, S. M., 1969. AcA 19, 245

\bibitem[Simon (1999)]{simon96} \u{S}imon, V., 1999. A\&AS 134, 1

\bibitem[van Hamme (1993)]{hamme93}
van Hamme, W., 1993. AJ 106, 2096

\bibitem[von Zeipel (1924)]{zeipel24}
von Zeipel, H., 1924. MNRAS 84, 665

\bibitem[Wilson and Devinney (1973)]{wilson73}
Wilson, R. E., Devinney, E. J., 1973. ApJ 82, 539

\bibitem[Wright (1936)]{wright36}
Wright, W.F., Dugan, R.S., 1936. AJ 45, 70

\bibitem[Zacharias et al.(2013)]{Zacharias13}
Zacharias, N., Finch, C.~T., Girard, T.~M., Henden, A., Bartlett, J.~L., Monet, D.~G.,
Zacharias, M.~I., 2013. AJ 145, 44
\end{thebibliography}
\end{document}